\documentclass[aps,prl,twocolumn,showpacs,floatfix,tightenlines,amsmath,amssymb,superscriptaddress,longbibliography]{revtex4-2}

\usepackage[hypertexnames=false]{hyperref}
\usepackage[version=4]{mhchem}
\usepackage[utf8]{inputenc}
\usepackage{graphicx}
\usepackage{xcolor}
\usepackage[normalem]{ulem}

\newcommand{\be}{\begin{equation}}
\newcommand{\ee}{\end{equation}}
\newcommand{\ba}{\begin{eqnarray}}
\newcommand{\ea}{\end{eqnarray}}

\begin{document}

\title{A Correlated Route to Antiferromagnetic Spintronics}

\author{Joel Bobadilla}
\affiliation{Universidad de Buenos Aires, Facultad de Ciencias Exactas y Naturales, Departamento de F\'isica, Buenos Aires, Argentina}

\author{Alberto Camjayi}
\email{alberto@df.uba.ar}
\affiliation{Universidad de Buenos Aires, Ciclo B\'asico Com\'un, Buenos Aires, Argentina}
\affiliation{CONICET--Universidad de Buenos Aires, Instituto de F\'isica de Buenos Aires (IFIBA), Buenos Aires, Argentina}

\date{\today}

\begin{abstract}
Antiferromagnets are attractive for spintronics owing to their vanishing net magnetization and ultrafast spin dynamics, yet their spin-compensated electronic structure has long confined them to passive roles. Here we identify a symmetry selection rule that overcomes this limitation: in a collinear antiferromagnet, a spin-polarized dc charge current requires the simultaneous breaking of particle--hole symmetry and of the equivalence between the two magnetic sublattices, either one alone leaving the polarization identically zero. We demonstrate the rule in the doped antiferromagnetic Hubbard model within dynamical mean-field theory: doping and a uniform magnetic field each break one of the two symmetries, and electronic correlations convert the resulting asymmetries into a sizable, field-tunable polarization of the charge current. The polarization is largest deep in the ordered phase and reverses sign at an emergent compensation point, set by the hierarchy of spin-dependent scattering rates and not dictated by symmetry, at which the dominant conducting spin species is interchanged. In correlated altermagnetic Hubbard models, the same two symmetries are broken structurally, by the sublattice-alternating hopping pattern: the field-driven mechanism identified here and the altermagnetic one are two realizations of a single selection rule. Electronic correlations thus emerge as an active ingredient for spintronics in structurally conventional collinear antiferromagnets.
%The same two symmetries, broken structurally by the crystal, underlie altermagnetic spin-polarized transport: the field-driven mechanism identified here and the altermagnetic one are two realizations of a single selection rule.%
\end{abstract}

\maketitle

\paragraph{Introduction}Spintronics exploits the electron spin, in addition to its charge, as an active degree of freedom in solid-state systems~\cite{Vedyaev2002,Inomata2008}. Harnessing this degree of freedom allows electrical currents to be manipulated at lower operating voltages, yielding reduced power consumption, faster operation, and higher data-storage density than conventional charge-based devices~\cite{Inomata2008}.

Ferromagnets (FMs) have long been the workhorse of spintronics: their broken time-reversal symmetry produces spin-split bands and a spontaneous magnetization, enabling early milestones such as spin-polarized tunneling and half-metallic conduction and making magnetic order easy to manipulate and read out~\cite{Vedyaev2002,Inomata2008}. That same spontaneous magnetization, however, is also a liability—stray fields limit integration density, the order is vulnerable to external magnetic perturbations, and the switching speed is capped by GHz-scale magnetization dynamics~\cite{Shim2025,Guo2025}.

Antiferromagnets (AFMs), with their fully compensated sublattices and vanishing net magnetization, sidestep these drawbacks: they are largely insensitive to external fields, support intrinsically faster, terahertz-scale spin dynamics, and permit denser device integration~\cite{Shim2025,Guo2025}. They are also materially abundant and diverse, ranging from insulators and semiconductors to metals and superconductors~\cite{Shim2025}.

Despite these appealing properties, conventional AFMs—composed of two magnetic sublattices with antiparallel moments and no additional local symmetry breaking—exhibit spin-degenerate electronic bands, enforced by the combined parity–time-reversal ($PT$) symmetry~\cite{Shim2025}. With no intrinsic spin splitting, anomalous and spin-polarized transport are suppressed, and such AFMs were long confined to passive roles in spintronic architectures~\cite{Guo2025}.

\begin{figure*}[t]
\centering
\includegraphics[width=1.0\textwidth]{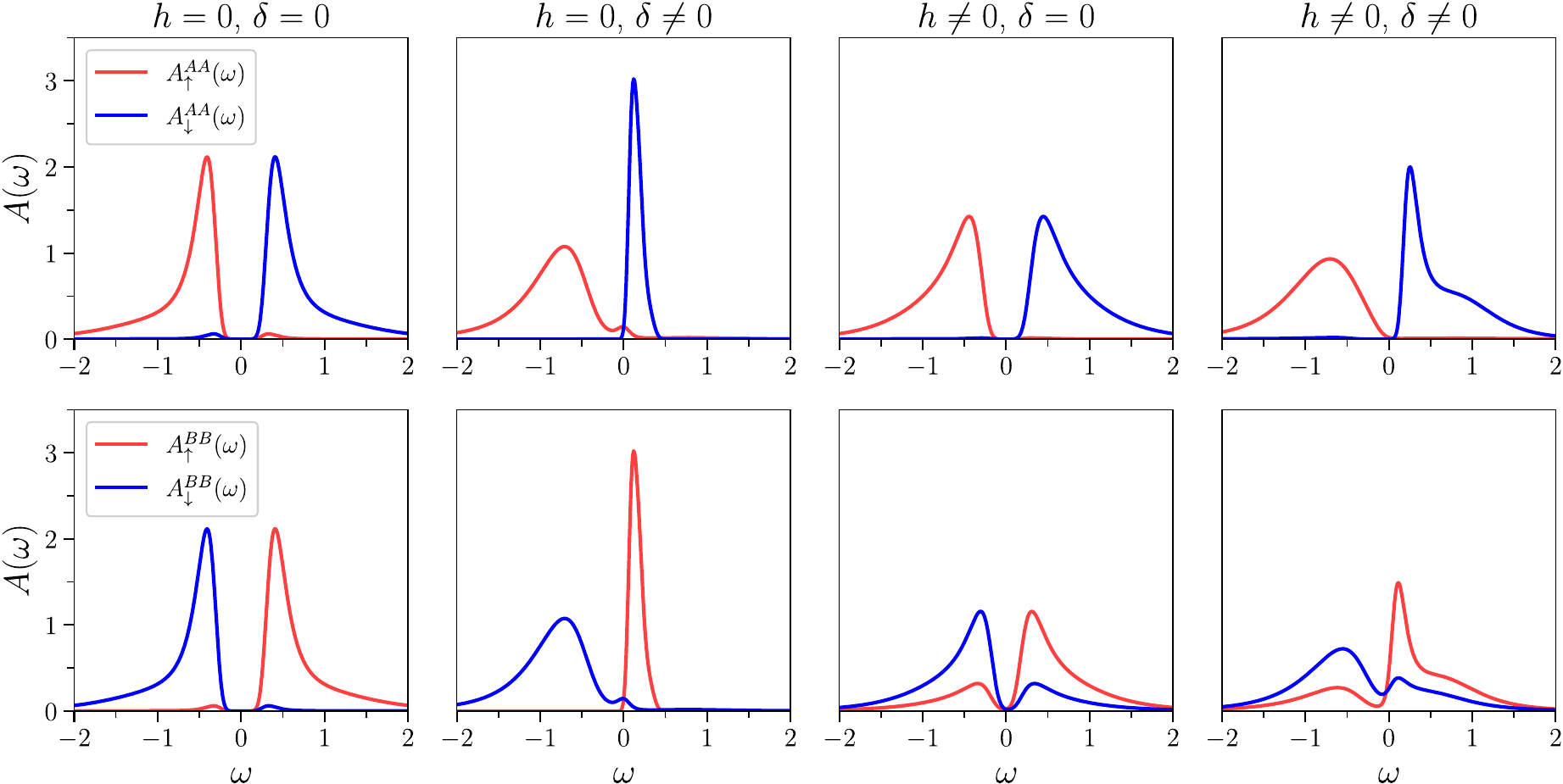}
\caption{%
Sublattice- and spin-resolved local spectral functions $A^{\alpha\alpha}_{\sigma}(\omega)$ for the four relevant regimes. From left to right: half filling without magnetic field, doped system at zero field, half-filled system under a magnetic field, and doped system under a magnetic field. Only the combined action of doping and magnetic field fully removes the spectral symmetries
enforcing spin-degenerate transport. Unless otherwise stated, the plots correspond to $h=0.07$ and $\delta=0.028$.
}
\label{fig:spectra}
\end{figure*}

This paradigm has been overturned by the discovery of spin-polarized AFMs, which can be classified into three broad families according to their distinct mechanisms for breaking parity–time-reversal ($PT$) symmetry: altermagnets (AMs), in which the two spin sublattices are related by crystal-rotation rather than translation or inversion symmetry; noncollinear AFMs, whose triangular or kagome spin textures break $PT$ symmetry and drive Berry-curvature–induced anomalous Hall responses; and layer-polarized AFMs, in which interlayer potential gradients enable electrically tunable, layer-selective spin polarization. As this family expands, a unified understanding of the symmetry principles governing their transport becomes increasingly important.

\begin{figure*}[t]
\centering
\includegraphics[width=\textwidth]{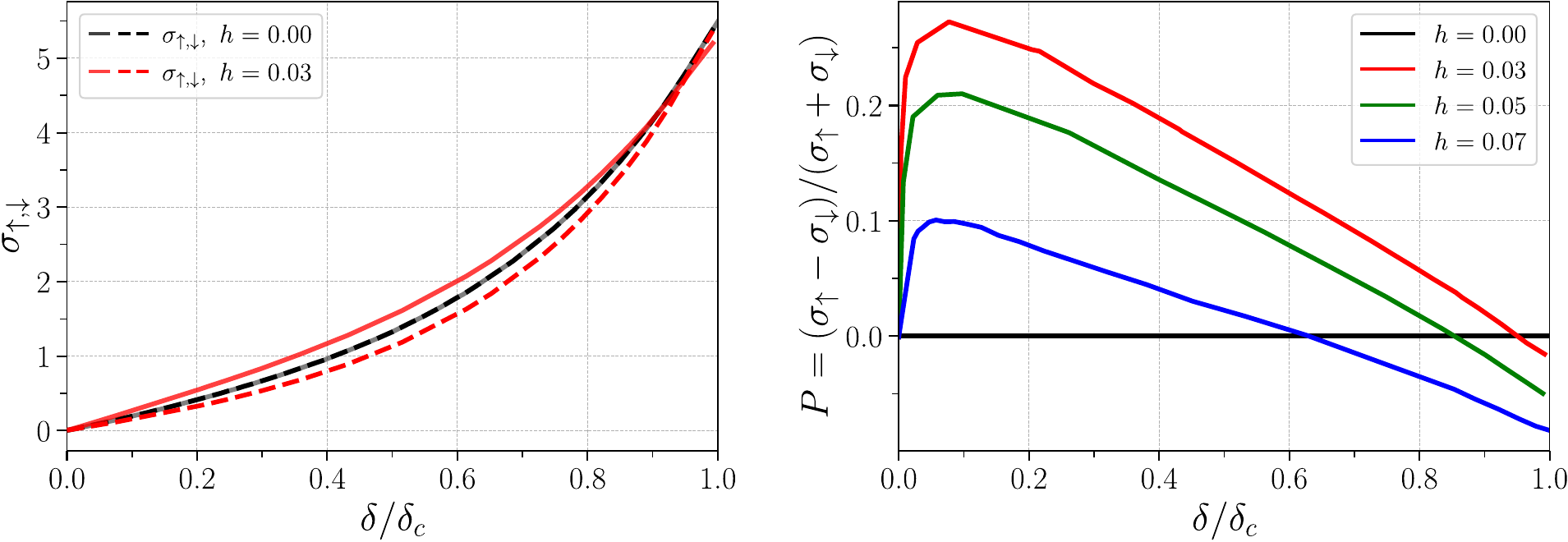}
\caption{%
Spin-resolved dc conductivities, $\sigma_{\uparrow}$, $\sigma_{\downarrow}$ (left) and current polarization, $P=(\sigma_{\uparrow}-\sigma_{\downarrow})/(\sigma_{\uparrow}+\sigma_{\downarrow})$ (right), as a function of doping within the antiferromagnetic regime, $\delta<\delta_c$, with $\delta_c$ the critical doping above which Néel order collapses. 
The left panel shows the spin-resolved dc conductivities for zero field ($h=0$) and for a representative finite field ($h=0.03$). The right panel shows the corresponding current polarization for several magnetic fields, highlighting the tunability of $P$ with doping and field strength. A finite spin polarization of the charge current emerges only when both particle--hole symmetry and sublattice equivalence are simultaneously broken.
}
\label{fig:transport}
\end{figure*}

Here we pursue a different route to the same end. Rather than engineering a crystal structure that breaks $PT$ symmetry, we ask whether a structurally conventional collinear AFM can be driven into spin-polarized transport by external means alone. To this end we study the Mott--Hubbard antiferromagnet---a paradigmatic correlated-electron model---under a uniform Zeeman field, solved within dynamical mean-field theory~\cite{Georges1996,Bobadilla2025}. Two external knobs each break one protecting symmetry: doping away from half filling breaks particle--hole ($PH$) symmetry, while the applied field lifts the dynamical equivalence of the two AFM sublattices associated with $PT$ symmetry. We find that neither breaking alone suffices: only when both act together are all spectral constraints relating opposite spins and sublattices removed (Fig.~\ref{fig:spectra}), and the dc charge current acquires a finite spin polarization (Fig.~\ref{fig:transport}).

\paragraph{Model and Methods} We study the one-band Hubbard model on a bipartite lattice in the presence of a uniform Zeeman field $h$,
\begin{multline}
H = -t\sum_{\langle ij \rangle, \sigma}\left( a^\dagger_{i \sigma} b_{j \sigma} + b^\dagger_{j \sigma} a_{i \sigma} \right)
+ U\sum_i n_{i\uparrow}n_{i\downarrow} \\
- \mu \sum_{i, \sigma} n_{i \sigma} - h \sum_{i} \left( n_{i\uparrow} - n_{i\downarrow} \right),
\label{H}
\end{multline}
where $a^\dagger$ ($b^\dagger$) creates an electron on sublattice $A$ ($B$), $U$ is the local Coulomb repulsion, and $\mu$ controls the electronic filling. We focus on the electron-doped regime, quantified by $\delta=n-1>0$, with $n$ the total filling per site, obtained by tuning $\mu>U/2$ away from half filling. We consider the infinite-dimensional bipartite hypercubic lattice, for which the noninteracting density of states is Gaussian and DMFT becomes exact.

To capture collinear antiferromagnetism, we solve the model within single-site DMFT by introducing two coupled impurity problems, one per sublattice~\cite{Bobadilla2025}. The impurity problems are solved using the continuous-time quantum Monte Carlo (CT-QMC) method~\cite{Haule2007}, yielding sublattice- and spin-resolved self-energies $\Sigma_{\alpha\sigma}(i\omega_n)$, where $\alpha=A,B$ labels the sublattice and $\sigma=\uparrow,\downarrow$ the spin projection. Real-frequency spectral functions $A^{\alpha\beta}_{\sigma}(\omega)$ are obtained by analytic continuation using the maximum entropy (MaxEnt) method~\cite{Levy2017}.

Throughout this work we fix the hopping amplitude to $t=0.5$, which sets the unit of energy, and choose $U=1.7$, placing the system in the intermediate-coupling regime. To avoid metamagnetic effects~\cite{Bobadilla2025,Held1997}, we work at a temperature $T=0.05$, corresponding to approximately $50\%$ of the Néel temperature ($T_N=0.092$), and vary the applied field $h$.

The spin-resolved dc conductivity $\sigma_{\sigma}$ is computed from the Kubo formula~\cite{Pruschke2003,Bobadilla2025}\footnote{Equation~(\ref{eq:sigmadc}) is the bubble contribution. In single-site DMFT, vertex corrections to the uniform dc conductivity vanish by the locality argument of Khurana~\cite{Khurana1990}: the self-energy and the irreducible two-particle vertex are local, so the odd momentum dependence of the current vertex cancels in the Brillouin-zone sum. This cancellation holds separately for each spin channel and is unaffected by a uniform Zeeman field~\cite{Vucicevic2021}.},
\begin{multline}
\sigma_{\sigma} = \sigma_0
\int \frac{d\omega}{2\pi} \int d\varepsilon \,
\left(-\frac{\partial f(\omega)}{\partial \omega}\right)
\rho(\varepsilon) \\
\times \left[
A^{AA}_{\sigma}(\varepsilon,\omega) A^{BB}_{\sigma}(\varepsilon,\omega)
+ \left(A^{AB}_{\sigma}(\varepsilon,\omega)\right)^2
\right],
\label{eq:sigmadc}
\end{multline}
where $\sigma_0$ is a dimensional constant~\footnote{The prefactor $\sigma_0$ fixes the physical units of the conductivity. It is proportional to $e^2/\hbar$ times the relevant lattice-length factor. Being independent of spin, doping, field, and interaction strength, $\sigma_0$ cancels in all dimensionless quantities reported here, in particular in the current polarization $P=(\sigma_\uparrow-\sigma_\downarrow)/(\sigma_\uparrow+\sigma_\downarrow)$.}, $f(\omega)$ is the Fermi--Dirac distribution, $\rho(\varepsilon)$ is the noninteracting density of states, and $A^{\alpha\beta}_{\sigma}(\varepsilon,\omega)$ are the interacting spectral functions resolved in spin and sublattice.

\paragraph{Results and Discussion}

Figure~\ref{fig:spectra} summarizes the sublattice- and spin-resolved local spectral functions in the four relevant regimes~\footnote{%
Within single-site DMFT on a bipartite lattice, the energy-resolved inter-sublattice spectral function $A^{AB}_{\sigma}(\varepsilon,\omega)$ is fully determined by the local self-energies $\Sigma^{A/B}_{\sigma}(\omega)$ and therefore inherits the same symmetry constraints. As a result, it does not introduce any additional spin asymmetry beyond that already encoded in the local spectral functions $A^{AA}_{\sigma}(\omega)$ and $A^{BB}_{\sigma}(\omega)$. A detailed discussion is provided in the Supplemental Material~\cite{SM}.%
}.
At half filling and zero field, the AFM state is characterized by the combined $PH$ and $PT$ symmetries, which enforce $A^{AA}_{\uparrow}(\omega)=A^{BB}_{\downarrow}(\omega)=A^{AA}_{\downarrow}(-\omega)=A^{BB}_{\uparrow}(-\omega)$. Because these local spectra follow from integrating the energy-resolved $A^{\alpha\beta}_{\sigma}(\varepsilon,\omega)$ over the noninteracting density of states, they inherit the symmetry constraints of the full interacting bands; the spin-resolved contributions to the Kubo kernel in Eq.~\eqref{eq:sigmadc} then remain strictly equal after the energy integration, yielding spin-degenerate dc transport.

Doping at zero field breaks $PH$ symmetry and drives strongly spin-dependent scattering, reflected in the asymmetric redistribution of spectral weight around the Fermi level. The equivalence of the two AFM sublattices, protected by $PT$, nonetheless survives—$A^{AA}_{\uparrow}(\omega)=A^{BB}_{\downarrow}(\omega)$ and $A^{AA}_{\downarrow}(\omega)=A^{BB}_{\uparrow}(\omega)$—and this residual symmetry still enforces exact compensation between spin channels in Eq.~\eqref{eq:sigmadc}, keeping the transport spin-degenerate with no net current polarization.

Applying a field at half filling instead lifts the sublattice equivalence and produces sublattice-dependent spectra, but $PH$ symmetry now imposes $A^{\alpha\alpha}_{\uparrow}(\omega)=A^{\alpha\alpha}_{\downarrow}(-\omega)$ on each sublattice $\alpha=A,B$. Since $-\partial_\omega f(\omega)$ is even in frequency, the two spin channels again contribute identically to Eq.~\eqref{eq:sigmadc}, and the transport stays spin-degenerate despite the explicit breaking of $PT$ symmetry.

Only with doping and field together are both $PH$ symmetry and the $PT$-protected sublattice equivalence removed. All spectral constraints linking opposite spins and sublattices are then lifted, the spin-resolved contributions to the Kubo kernel no longer compensate, and the charge current acquires a finite spin polarization—the selection rule anticipated above.

Figure~\ref{fig:transport} shows how these considerations play out in transport. The left panel gives the spin-resolved conductivities versus doping, at zero field and at a representative $h=0.03$. At zero field they remain strictly degenerate at all dopings, $\sigma_\uparrow=\sigma_\downarrow$, even though doping has induced strong spin-dependent scattering—confirming that breaking $PH$ symmetry alone cannot polarize the current.

A finite field splits the conductivities, and the current polarization $P=(\sigma_\uparrow-\sigma_\downarrow)/(\sigma_\uparrow+\sigma_\downarrow)$ (right panel) becomes nonzero only when doping and field act together, i.e., when $PH$ symmetry and the $PT$-protected sublattice equivalence are broken simultaneously.

The polarization varies strongly and non-monotonically with doping. From $P=0$ at half filling it rises steeply to a maximum at low doping—where long-range AFM order is most robust, and already at modest fields~\footnote{The Zeeman term in Eq.~(\ref{H}) couples $h$ to $n_{i\uparrow}-n_{i\downarrow} = 2S^z_i$, so that $h = g\mu_B B/2$, i.e., $h = \mu_B B$ for $g = 2$. Since the transport response is organized by the ratio of $h$ to the ordering scale, the natural conversion anchors the N\'eel temperature: $\mu_B B = (h/T_N)\,k_B T_N^{\mathrm{exp}}$. For a material with $T_N^{\mathrm{exp}} = 100$~K, the representative field $h = 0.03 \simeq 0.33\,T_N$ corresponds to $B \simeq 49$~T; these fields scale linearly with $T_N^{\mathrm{exp}}$, dropping to the ten-tesla range for antiferromagnets with $T_N^{\mathrm{exp}} \lesssim 30$~K. No threshold field is involved: at finite doping the polarization grows continuously from zero with the applied field, which sets its magnitude rather than its existence.}—then falls and reverses sign at an intermediate doping, turning negative before the Néel order collapses at $\delta_c$ (right panel; Supplemental Material~\cite{SM}).

This sign reversal marks an emergent compensation point. Unlike the vanishing of $P$ at half filling or at zero field, it is not protected by symmetry: it arises from the balance of two competing correlation effects that jointly set the spin-resolved conductivities---the spin imbalance of the spectral weight at the Fermi level and the hierarchy of spin- and sublattice-resolved scattering rates $\tau^{-1}_{\alpha\sigma}=-2\,\mathrm{Im}\,\Sigma_{\alpha\sigma}(\omega\to0^{+})$ (Supplemental Material~\cite{SM}). At low doping both effects favor the same spin channel; as the staggered order weakens with doping, spectral weight is redistributed and the relative lifetimes of the two channels evolve, until $\sigma_\uparrow=\sigma_\downarrow$. Beyond this compensation point the sign of the polarization reverses: the spin channel that carries the larger share of the charge current hands over to the other, so that doping alone interchanges the dominant conducting species within a single antiferromagnetic phase---a correlation-driven transport crossover with no counterpart in the symmetry analysis. A stronger field shifts the compensation point to lower doping and steadily narrows the positive-polarization window. That $P$ is largest at low doping, deep in the ordered phase where N\'eel order is most robust, and is already sizable at modest fields, further indicates that the effect is an intrinsic property of the correlated antiferromagnet rather than a vestige of proximity to a paramagnetic metal: the paramagnetic solution of the same model yields only a weak polarization of opposite sign, as shown in the Supplemental Material~\cite{SM}.

\paragraph{Analogy with Altermagnets}
The results above distill into a symmetry selection rule: in a collinear antiferromagnet, a spin-polarized dc charge current requires the simultaneous breaking of $PH$ symmetry and of the equivalence between the two magnetic sublattices, with either one alone leaving the polarization identically zero. We now argue that the same rule underlies the spin-polarized transport of correlated altermagnetic Hubbard models~\cite{Das2024,He2025,DelRe2025}, so that the connection between the two phenomena is a shared symmetry requirement. In altermagnets, spin-split bands and currents arise in collinear, fully compensated magnets because the opposite-spin sublattices are related by crystal rotations rather than by translation or inversion, breaking the combined parity--time-reversal ($PT$) symmetry~\cite{Shim2025,Guo2025}; closely related spin-resolved responses have been studied in Hubbard models engineered to host altermagnetic order through modified hopping patterns~\cite{Das2024,He2025,DelRe2025}.

% We now argue that this is the very rule underlying altermagnetic spin-polarized transport, so that the connection between the two phenomena is a shared symmetry requirement

The correspondence becomes explicit in the $t$--$t'$--$\delta$ models of Refs.~\cite{He2025,DelRe2025}, where the intra-sublattice dispersions read
\begin{multline}
f_{A/B}(\mathbf{k})
=
-2t'\left(\cos k_x+\cos k_y\right)
\\[3pt]
\mp 2t''\left(\cos k_x-\cos k_y\right),
\label{eq:AMdispersion}
\end{multline}
with $t'$ the isotropic and $t''$ the sublattice-alternating next-nearest-neighbor hopping (the latter written $t'\delta$ in Refs.~\cite{He2025,DelRe2025}, a symbol we reserve for the doping). The isotropic part, identical on both sublattices, is an intra-sublattice hopping and as such breaks $PH$ symmetry. The anisotropic part, of opposite sign on $A$ and $B$, breaks the sublattice equivalence: the two sublattices remain related only by a rotation combined with time reversal, and $PT$ symmetry is lost. Neither breaking alone yields spin-split transport---at $t''=0$ the model is a $PH$-broken but sublattice-symmetric AFM with spin-degenerate transport---in one-to-one correspondence with our selection rule, the structural couplings $(t',t'')$ assuming the roles played here by doping and field. The mapping also explains an apparent asymmetry between the two settings: because $t'$ breaks $PH$ symmetry already at half filling, the altermagnet spin-splits without doping, whereas our nearest-neighbor AFM---$PH$-symmetric at half filling---must be doped.

The two realizations are not phenomenologically identical, and the differences are physically meaningful. In the altermagnet both symmetries are broken structurally: the spin splitting is spontaneous, occurs at zero field with strictly zero net magnetization, and resides in the band dispersions as a momentum-anisotropic, $d$-wave feature. The residual rotation--time-reversal symmetry ties the two spin channels to orthogonal directions, $\sigma^{xx}_\uparrow=\sigma^{yy}_\downarrow$: the current polarization is finite along a generic axis but reverses sign under a $90^{\circ}$ rotation---a $d$-wave polarization that averages to zero over directions---while the local self-energies remain locked, $\Sigma_{A\uparrow}=\Sigma_{B\downarrow}$. In our conventional AFM the lattice preserves $PT$ and the isotropic geometry offers no such directional channel; the sublattice equivalence can be lifted only through the spin sector, by the field, which breaks that local locking. The spin asymmetry in the sublattice-resolved spectral weight and scattering rates then yields a net isotropic current polarization that survives directional averaging. The small uniform magnetization this entails is not the origin of the effect but the order parameter conjugate to the symmetry-breaking field, marking a field-driven rather than structural realization of the same rule.

Within this class of models, structural and field-driven spin-polarized transport thus emerge as two realizations of a single symmetry principle---one written permanently into the hopping pattern, the other switched on, and continuously tuned, by doping and applied field. Spin-polarized charge transport in compensated collinear magnets is therefore not the privilege of special crystal classes: it is available, on demand, in the broad family of structurally conventional correlated antiferromagnets.

% Altermagnetic and field-driven spin-polarized transport thus emerge as two realizations of a single symmetry principle---one written permanently into the crystal structure, the other switched on, and continuously tuned, by doping and applied field.

\paragraph{Conclusions}
We have established a symmetry selection rule for spin-polarized charge transport in collinear antiferromagnets and demonstrated its field-driven realization in the doped antiferromagnetic Hubbard model. The rule fixes when the polarization may be nonzero; the correlations fix everything quantitative---its magnitude; its non-monotonic doping dependence; and its sign, which reverses at an emergent compensation point, set by the hierarchy of spin- and sublattice-resolved scattering rates, where doping and field interchange the dominant conducting spin species without reorienting the N\'eel order. Whether correlation-driven spin-dependent scattering also shapes transport in altermagnetic materials---where particle--hole symmetry is absent from the outset---is an open question. Beyond the one-band model, multiorbital and material-specific realizations---where crystal-field splitting and Hund coupling should further enrich the scattering hierarchy---are a natural next step, as is the experimental search for the onset of current polarization in lightly doped antiferromagnetic Mott insulators under applied magnetic fields.

% Bibliografía
%\bibliographystyle{apsrev4-2}
\bibliography{referencias}

\end{document}

\title{Supplemental Material for\\
A Correlated Route to Antiferromagnetic Spintronics}

\author{Joel Bobadilla}
\author{Alberto Camjayi}

\maketitle

\section{Symmetry analysis of the bipartite Hubbard model under a uniform magnetic field}
\label{SM:Symmetry}

In this section we analyze the symmetries of the bipartite Hubbard model in the presence of a uniform magnetic field, given by Eq.~(1) of the main text,
\begin{equation}
H = -t\sum_{\langle ij \rangle, \sigma}
\left(
a^\dagger_{i \sigma} b_{j \sigma}
+ b^\dagger_{j \sigma} a_{i \sigma}
\right)
+ U\sum_i n_{i\uparrow}n_{i\downarrow} 
- \mu \sum_{i, \sigma} n_{i \sigma}
- h \sum_{i} \left( n_{i\uparrow} - n_{i\downarrow} \right) ,
\label{eq:H_SM}
\end{equation}
where $a^\dagger_{i\sigma}$ ($b^\dagger_{i\sigma}$) creates an electron with spin $\sigma$ on sublattice $A$ ($B$), $U$ is the local Coulomb repulsion, $\mu$ controls the filling, and $h$ denotes a uniform Zeeman-like magnetic field.

\subsection{Particle--hole symmetry on a bipartite lattice}
\label{SM:PH}

On a bipartite lattice composed of two sublattices $A$ and $B$, the particle--hole (PH) transformation is defined as,
\begin{equation}
c_{i\sigma} \;\longrightarrow\; \eta_i\, c_{i\sigma}^\dagger,
\qquad
\eta_i =
\begin{cases}
+1, & i\in A,\\
-1, & i\in B.
\end{cases}
\label{eq:PH}
\end{equation}
At half filling ($\mu = U/2$), the Hubbard model in the absence of external fields is invariant under this transformation~\cite{Arovas2022}. 

The inclusion of a uniform magnetic field via the Zeeman term, $-h\sum_i (n_{i\uparrow}-n_{i\downarrow})$, however, explicitly breaks PH symmetry. Under the transformation~\eqref{eq:PH}, the local density operators transform as,
\begin{equation}
n_{i\sigma} \;\longrightarrow\; 1 - n_{i\sigma},
\end{equation}
so that the local magnetization changes sign,
\begin{equation}
n_{i\uparrow} - n_{i\downarrow} \;\longrightarrow\; -\left(n_{i\uparrow} - n_{i\downarrow}\right).
\end{equation}
As a result, the Zeeman term changes sign under PH, and the Hubbard Hamiltonian in the presence of a uniform magnetic field does not possess particle--hole symmetry, even at half filling.

\subsection{Combined particle--hole and spin-flip symmetry}
\label{SM:PHsp}

Although pure particle--hole symmetry is broken by the Zeeman term, the bipartite Hubbard Hamiltonian at half filling remains invariant under a combined transformation consisting of particle--hole conjugation and a spin flip, even in the presence of a uniform magnetic field. We denote this symmetry as PHsp.

The PHsp transformation acts on the fermionic operators as,
\begin{equation}
\label{eq:PHsp}
\begin{aligned}
c_{i\uparrow} &\longrightarrow \eta_i\, c_{i\downarrow}^\dagger,\\
c_{i\downarrow} &\longrightarrow -\eta_i\, c_{i\uparrow}^\dagger,
\end{aligned}
\end{equation}
where $\eta_i=+1$ ($-1$) for sites belonging to sublattice $A$ ($B$). The relative minus sign ensures that the transformation is canonical and preserves fermionic anticommutation relations.

Under PHsp, the local density operators transform as,
\begin{equation}
n_{i\uparrow} \longrightarrow 1 - n_{i\downarrow},
\qquad
n_{i\downarrow} \longrightarrow 1 - n_{i\uparrow},
\end{equation}
which implies that the local magnetization remains invariant,
\begin{equation}
n_{i\uparrow} - n_{i\downarrow} \;\longrightarrow\; n_{i\uparrow} - n_{i\downarrow}.
\end{equation}
As a consequence, the Zeeman term is invariant under PHsp. Together with the condition $\mu = U/2$, this establishes PHsp as an exact symmetry of the half-filled Hubbard model in the presence of a uniform magnetic field.

\subsection{Consequences for Green functions and spectral functions}
\label{SM:GreenSpectral}

We now derive the implications of PHsp symmetry for single-particle Green functions and spectral functions. We consider the retarded Green function resolved in sublattice, spin, and band energy,
\begin{equation}
G^{\alpha\beta}_\sigma(\varepsilon,t)
=
- i \theta(t)
\left\langle
\left\{
c_{\alpha\sigma}(\varepsilon,t),
c^\dagger_{\beta\sigma}(\varepsilon,0)
\right\}
\right\rangle,
\end{equation}
where $\alpha,\beta=A,B$ label the sublattices and $\sigma=\uparrow,\downarrow$.

At half filling, the thermal average is invariant under the PHsp transformation. Applying Eq.~\eqref{eq:PHsp} to the fermionic operators and using the symmetry properties of the anticommutator, we obtain,
\begin{equation}
\label{eq:GPHsp_t}
G^{\alpha\beta}_\sigma(\varepsilon,t)
=
-\,\eta_\alpha \eta_\beta
\left[
G^{\alpha\beta}_{\bar\sigma}(\varepsilon,t)
\right]^*,
\end{equation}
where $\bar\sigma$ denotes the opposite spin projection. Fourier transforming to frequency space yields,
\begin{equation}
\label{eq:GPHsp_w}
G^{\alpha\beta}_\sigma(\varepsilon,\omega)
=
-\,\eta_\alpha \eta_\beta
\left[
G^{\alpha\beta}_{\bar\sigma}(\varepsilon,-\omega)
\right]^*.
\end{equation}

The interacting spectral functions are defined as,
\begin{equation}
A^{\alpha\beta}_\sigma(\varepsilon,\omega)
=
-\frac{1}{\pi}\,
\mathrm{Im}\,
G^{\alpha\beta}_\sigma(\varepsilon,\omega).
\end{equation}
Using Eq.~\eqref{eq:GPHsp_w}, we obtain the central symmetry relation,
\begin{equation}
\label{eq:APHsp}
A^{\alpha\beta}_\sigma(\varepsilon,\omega)
=
\eta_\alpha \eta_\beta\,
A^{\alpha\beta}_{\bar\sigma}(\varepsilon,-\omega).
\end{equation}

In particular, this implies the following relations for the diagonal and off-diagonal components,
\begin{align}
A^{AA}_\uparrow(\varepsilon,\omega)
&=
A^{AA}_\downarrow(\varepsilon,-\omega),
&
A^{BB}_\uparrow(\varepsilon,\omega)
&=
A^{BB}_\downarrow(\varepsilon,-\omega),
&
A^{AB}_\uparrow(\varepsilon,\omega)
&=
- A^{AB}_\downarrow(\varepsilon,-\omega).
\end{align}

These relations hold exactly at half filling in the presence of a uniform magnetic field and form the basis for the spin-degenerate dc transport response discussed in the main text.

We now use these exact symmetry relations to analyze their consequences for the dc conductivity kernel entering the Kubo formula. In particular, we show how the symmetry properties of the energy-resolved spectral functions lead to an exact compensation between spin channels in the dc transport response whenever generalized particle--hole symmetry (pure PH at zero magnetic field or PHsp in the presence of a magnetic field) or sublattice equivalence is preserved.\footnote{In the following, we use the shorthand ``PH symmetry'' to collectively refer to pure particle--hole symmetry at zero magnetic field and to the combined particle--hole and spin-flip (PHsp) symmetry at half filling in the presence of a magnetic field. This notational simplification is adopted for clarity and does not affect the generality of the arguments.}

\section{Implications for the dc conductivity kernel}
\label{SM:Kubo}

We now analyze how the symmetry relations derived in the previous section translate into constraints on the dc conductivity computed from the Kubo formula. The spin-resolved dc conductivity reads, 
\begin{equation}
\sigma_{\sigma} = \sigma_0
\int \frac{d\omega}{2\pi} \int d\varepsilon \,
\left(-\frac{\partial f(\omega)}{\partial \omega}\right)
\rho(\varepsilon)
\left[
A^{AA}_{\sigma}(\varepsilon,\omega) A^{BB}_{\sigma}(\varepsilon,\omega)
+ \left(A^{AB}_{\sigma}(\varepsilon,\omega)\right)^2
\right],
\label{eq:Kubo_SM}
\end{equation}
where $\sigma_0$ is a dimensional constant, $\rho(\varepsilon)$ is the noninteracting density of states and $f(\omega)$ is the Fermi--Dirac distribution.

\begin{figure}[t]
    \centering
    \includegraphics[width=\linewidth]{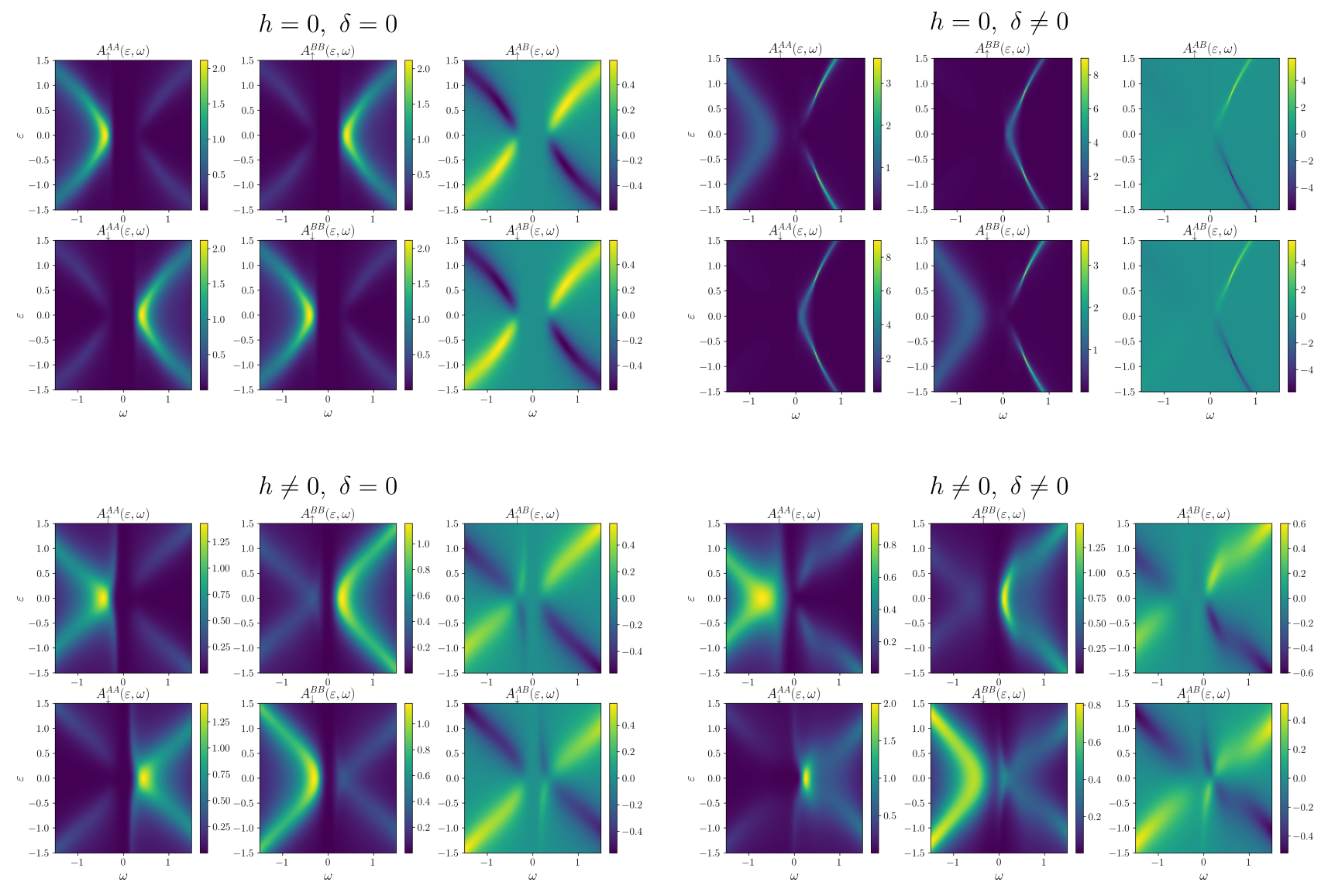}
    \caption{Energy-resolved spectral functions $A^{\alpha\beta}_{\uparrow}(\varepsilon,\omega)$ entering the dc conductivity kernel for the four regimes discussed in the main text. Each block corresponds to one column of Fig.~1 of the main text: (top left) $h=0$, $\delta=0$; (top right) $h=0$, $\delta\neq0$; (bottom left) $h\neq0$, $\delta=0$; (bottom right) $h\neq0$, $\delta\neq0$. Within each block, the panels display the diagonal ($A^{AA}$, $A^{BB}$) and off-diagonal ($A^{AB}$) spectral functions resolved in band energy $\varepsilon$ and frequency $\omega$. For $h=0$ and $\delta=0$, both particle--hole symmetry and antiferromagnetic sublattice equivalence are preserved, enforcing reflection symmetry with respect to $\omega=0$ and exact compensation between spin channels in the Kubo kernel. At zero field but finite doping ($h=0$, $\delta\neq0$), particle--hole symmetry is broken and the spectral functions lose their reflection symmetry in $\omega$, yet the equivalence between antiferromagnetic sublattices remains intact, ensuring spin-degenerate dc transport. For finite magnetic field at half filling ($h\neq0$, $\delta=0$), sublattice equivalence is lifted, but the generalized particle--hole (PHsp) symmetry enforces the relations $A^{\alpha\beta}_{\uparrow}(\varepsilon,\omega)=\eta_\alpha\eta_\beta A^{\alpha\beta}_{\downarrow}(\varepsilon,-\omega)$, which, together with the even parity of $-\partial_\omega f(\omega)$, again leads to exact cancellation between spin channels. Only when both finite doping and magnetic field are present ($h\neq0$, $\delta\neq0$) are all symmetry constraints removed, resulting in uncompensated contributions to the dc conductivity kernel and a finite spin polarization of the charge current.}
    \label{fig:SM_spectral_maps}
\end{figure}

The symmetry properties of the kernel in Eq.~\eqref{eq:Kubo_SM} play a central role. In particular, the derivative $-\partial_\omega f(\omega)$ is an even function of frequency, so that any relation between spectral functions at opposite frequencies directly constrains the spin-resolved contributions to the conductivity. These constraints are illustrated explicitly in Fig.~\ref{fig:SM_spectral_maps} through the energy-resolved spectral functions entering the Kubo kernel in the four regimes discussed in the main text.

At half filling, generalized particle--hole symmetry implies the relations,
\begin{equation}
A^{\alpha\beta}_\uparrow(\varepsilon,\omega)
=
\eta_\alpha \eta_\beta\,
A^{\alpha\beta}_\downarrow(\varepsilon,-\omega).
\end{equation}
Inserting these relations into the kernel of Eq.~\eqref{eq:Kubo_SM} and using the even parity of $-\partial_\omega f(\omega)$, one finds that the spin-resolved contributions to the dc conductivity are exactly equal,
\begin{equation}
\sigma_\uparrow = \sigma_\downarrow,
\end{equation}
despite the presence of a finite magnetic field and sublattice-dependent spectral functions.

Away from half filling, particle--hole symmetry is broken and the spectral functions generally lose their reflection symmetry with respect to $\omega=0$. Nevertheless, when the sublattice equivalence protected by $PT$ symmetry is preserved (i.e., at zero magnetic field), the relations
\begin{equation}
A^{AA}_{\uparrow}(\varepsilon,\omega)
=
A^{BB}_{\downarrow}(\varepsilon,\omega),
\qquad
A^{AA}_{\downarrow}(\varepsilon,\omega)
=
A^{BB}_{\uparrow}(\varepsilon,\omega),
\qquad
A^{AB}_{\uparrow}(\varepsilon,\omega)
=
A^{AB}_{\downarrow}(\varepsilon,\omega).
\end{equation}
ensure an exact compensation between spin channels in the conductivity kernel, again yielding spin-degenerate dc transport.

Only when both generalized particle--hole symmetry and sublattice equivalence are simultaneously broken—namely, at finite doping and in the presence of a magnetic field—do the spin-resolved contributions to the Kubo kernel fail to compensate. In this regime, the dc conductivity becomes spin dependent, giving rise to a finite spin polarization of the charge current, as discussed in the main text.

\section{Magnetic phase diagram}
\label{SM:PhaseDiagram}

\begin{figure}[t]
\centering
\includegraphics[width=0.8\linewidth]{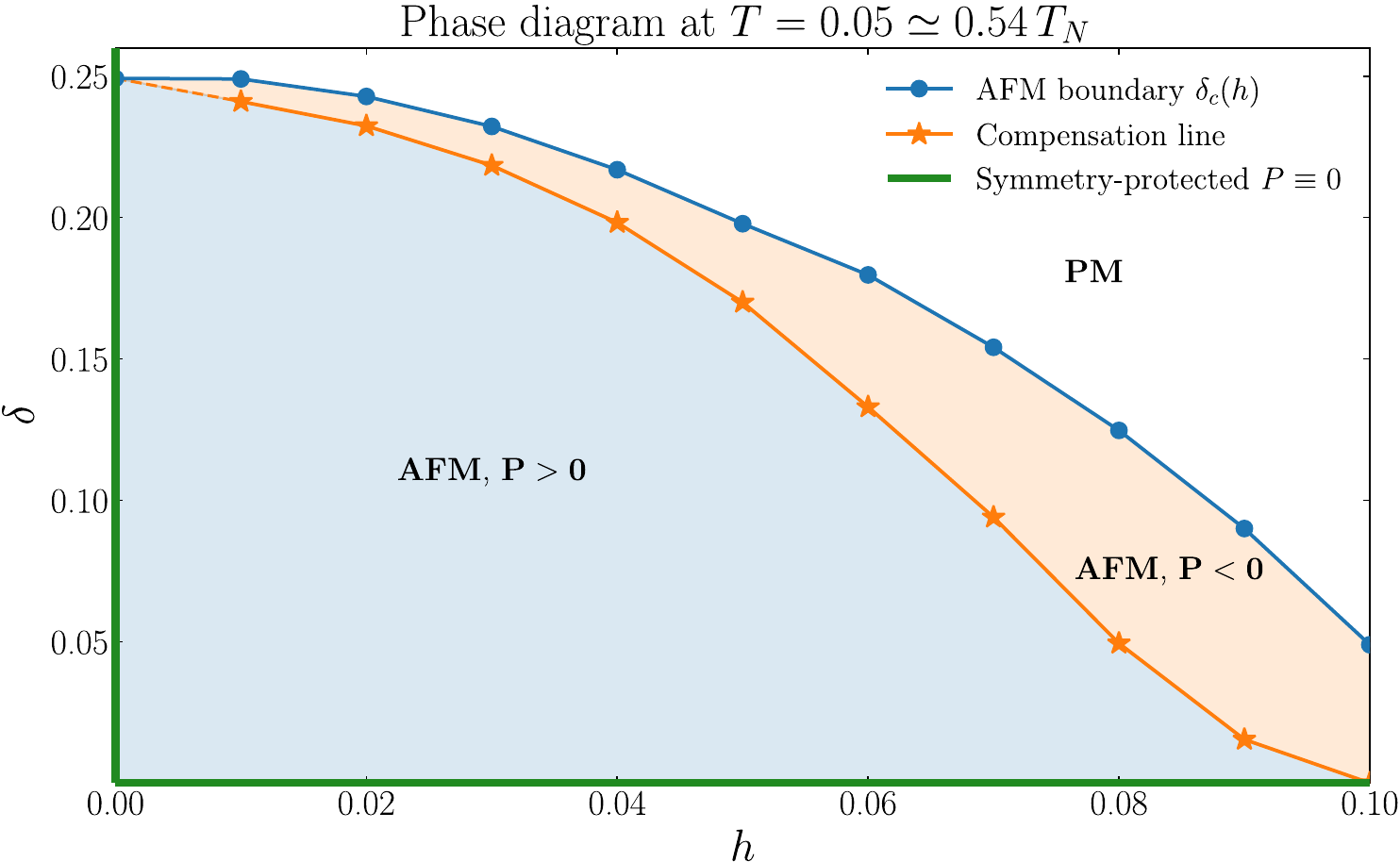}
\caption{Magnetic phase diagram in the $(h,\delta)$ plane at $T = 0.05 \simeq 0.54\,T_N$. Blue circles: antiferromagnetic phase boundary $\delta_c(h)$, separating the N\'eel-ordered region from the paramagnet (PM); the boundary shrinks monotonically with field. Orange stars: compensation line, the locus at which the spin-resolved conductivities cross, $\sigma_\uparrow = \sigma_\downarrow$, and the current polarization changes sign; the dashed segment indicates its $h \to 0^{+}$ limit, where the line is pushed onto the phase boundary. Green lines: symmetry-protected zeros of the polarization---along $\delta = 0$, generalized particle--hole symmetry enforces $P \equiv 0$ at any field, and along $h = 0$, the sublattice equivalence enforces $P \equiv 0$ at any doping. The compensation line, by contrast, is an emergent zero of the correlated state, not protected by symmetry. The two data curves partition the ordered phase into a region of positive polarization at low doping (blue shading) and a wedge of negative polarization adjacent to the phase boundary (orange shading). All calculations lie within the ordered region; lines connecting symbols are guides to the eye.}
\label{fig:SM_phase_diagram}
\end{figure}

The N\'eel temperature of the model at half filling and zero field is $T_N = 0.092$ in the units of the main text. Doping suppresses the staggered order: at $h=0$, $T_N$ decreases monotonically with $\delta$, and at the working temperature $T = 0.05 \simeq 0.54\,T_N$ the N\'eel state survives up to a critical doping $\delta_c \simeq 0.25$, beyond which it collapses. A uniform magnetic field competes with the staggered order, and the phase boundary $\delta_c(h)$ shrinks monotonically with increasing field, as shown in Fig.~\ref{fig:SM_phase_diagram}. All results reported in the main text and in this Supplemental Material lie within the ordered phase, $\delta < \delta_c(h)$; accordingly, dopings in the transport figures are shown normalized to the corresponding $\delta_c(h)$.

The phase diagram also locates the compensation line: the locus of points at which the spin-resolved conductivities cross, $\sigma_\uparrow = \sigma_\downarrow$, and the current polarization changes sign. It partitions the ordered phase into a low-doping region of positive polarization and a wedge of negative polarization adjacent to the phase boundary. As the field increases, the compensation line moves to lower doping while the boundary itself contracts, so the positive-polarization region shrinks from both sides; by $h \simeq 0.1$ the compensation line reaches vanishing doping and the positive-polarization window closes. The two axes of the diagram play a distinguished role: along $\delta = 0$ the generalized particle--hole symmetry enforces $P \equiv 0$ at any field, and along $h = 0$ the sublattice equivalence enforces $P \equiv 0$ at any doping (green lines in Fig.~\ref{fig:SM_phase_diagram}). These are symmetry-protected zeros, qualitatively distinct from the emergent zero traced by the compensation line; consistently, as $h \to 0^{+}$ the compensation line is pushed onto the phase boundary and the negative-polarization wedge closes, in agreement with the trend of the sign-change dopings in Fig.~\ref{fig:SM_polarization}. The microscopic origin of the compensation line is analyzed in the following Section.

\section{Scattering-rate hierarchy and current polarization}
\label{SM:Scattering}

\begin{figure}[t]
\centering
\includegraphics[width=\linewidth]{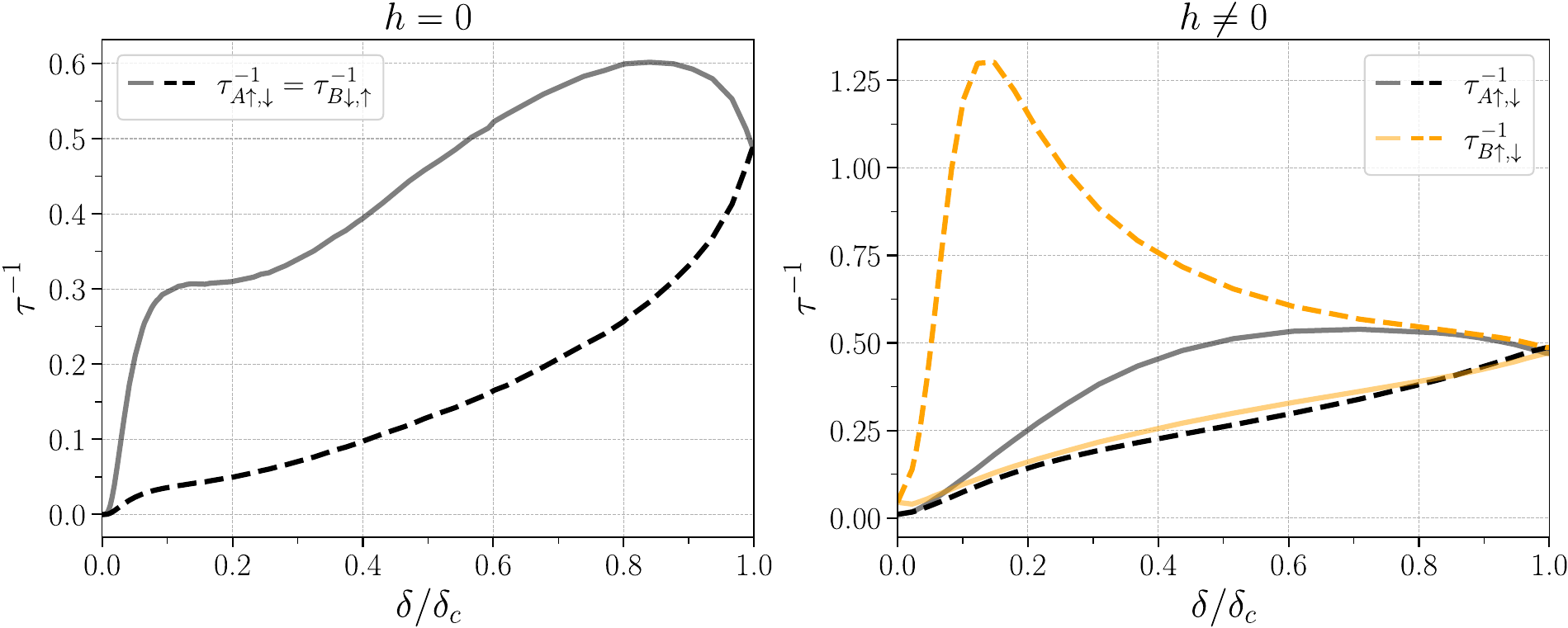}
\caption{Spin- and sublattice-resolved scattering rates $\tau^{-1}_{\alpha\sigma}$ as a function of normalized doping $\delta/\delta_c$. Left panel: zero magnetic field ($h=0$). Doping alone induces a clear spin-dependent splitting of the scattering rates, reflecting the breaking of particle--hole symmetry while preserving antiferromagnetic sublattice equivalence. Right panel: finite magnetic field ($h=0.07$). In this regime, both spin and sublattice splittings are present, leading to a hierarchy of scattering rates that directly feeds into the uncompensated contributions of the Kubo kernel.}
\label{fig:SM_scattering}
\end{figure}

To further elucidate the microscopic mechanism, we analyze the evolution of the spin- and sublattice-resolved scattering rates\footnote{The scattering rates are defined from the imaginary part of the local self-energy at the Fermi level, $\tau^{-1}_{\alpha\sigma} = -2\,\mathrm{Im}\,\Sigma_{\alpha\sigma}(\omega \to 0^{+})$, with $\alpha=A,B$ the sublattice and $\sigma=\uparrow,\downarrow$ the spin projection. The real-frequency self-energies follow from maximum-entropy analytic continuation of the CT-QMC Matsubara data, and the limit $\omega \to 0^{+}$ is taken on $-\mathrm{Im}\,\Sigma_{\alpha\sigma}(\omega)$. No quasiparticle-weight renormalization is applied: $\tau^{-1}_{\alpha\sigma}$ is the bare broadening entering the spectral functions, not the quasiparticle inverse lifetime $Z_{\alpha\sigma}\,\tau^{-1}_{\alpha\sigma}$.} as a function of doping. As shown in Fig.~\ref{fig:SM_scattering}, doping alone already induces a pronounced spin asymmetry in the scattering rates, even in the absence of a magnetic field. However, because sublattice equivalence remains intact at $h=0$, the Kubo kernel still enforces exact spin compensation. When a finite magnetic field is applied, sublattice symmetry is lifted and the resulting hierarchy of $\tau^{-1}_{\alpha\sigma}$ removes this compensation, allowing spin-dependent transport to emerge.

\begin{figure}[t]
\centering
\includegraphics[width=0.7\linewidth]{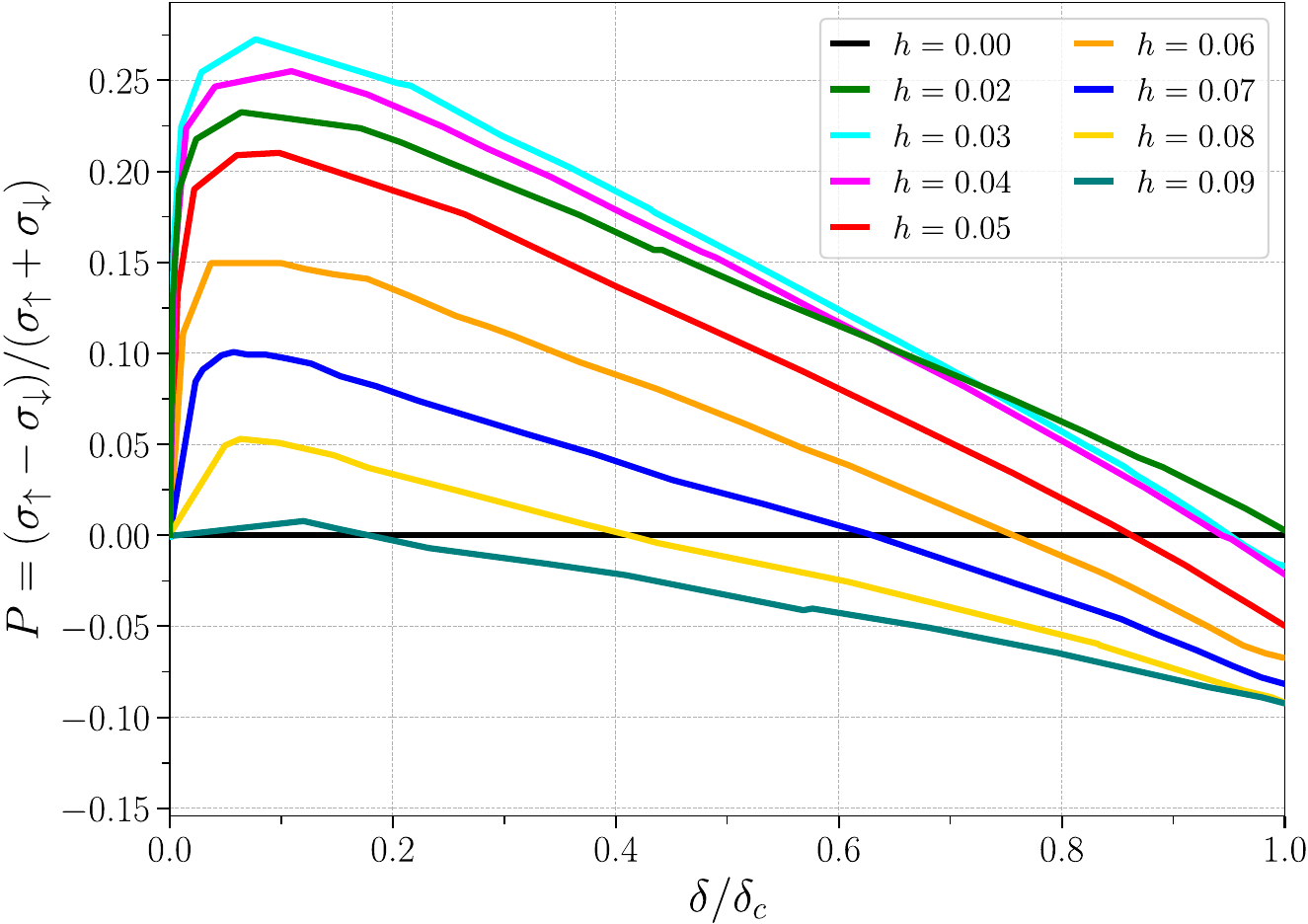}
\caption{Spin polarization of the dc charge current,
$P=(\sigma_\uparrow-\sigma_\downarrow)/(\sigma_\uparrow+\sigma_\downarrow)$,
as a function of normalized doping $\delta/\delta_c$ for different values of the magnetic field $h$. At zero field, the polarization vanishes identically for all dopings due to exact spin compensation in the Kubo kernel. Finite magnetic fields progressively lift this compensation, producing a sizable and tunable spin polarization. The sign reversal reflects the interplay between spin-dependent scattering and spectral weight redistribution, demonstrating the field-controlled character of the correlated spintronic response.}
\label{fig:SM_polarization}
\end{figure}

Under a finite Zeeman field ($h=0.07$ in the right panel of Fig.~\ref{fig:SM_scattering}), the low-doping regime exhibits a pronounced asymmetry between $\tau^{-1}_{A\uparrow}$ and $\tau^{-1}_{B\downarrow}$, whereas $\tau^{-1}_{A\downarrow}$ and $\tau^{-1}_{B\uparrow}$ remain small and nearly equal. As the doping approaches $\delta_c$ and sublattice equivalence is progressively restored, the four scattering rates converge; in the intermediate range $0.6\lesssim\delta/\delta_c\lesssim0.8$, $\tau^{-1}_{A\uparrow}$ and $\tau^{-1}_{B\downarrow}$ become comparable. Together with the concomitant redistribution of spectral weight, this evolution produces the compensation point at which $\sigma_\uparrow=\sigma_\downarrow$ and the dominant conducting spin species is interchanged, consistent with the location of the sign reversal of $P$ for this field in Fig.~\ref{fig:SM_polarization}.

The impact of this hierarchy of scattering rates on transport is directly reflected in the spin polarization of the dc current. Figure~\ref{fig:SM_polarization} highlights the strong field tunability of the correlated spintronic response. The polarization magnitude can be continuously controlled by the external magnetic field. The sign reversal reflects the competition between enhanced spin-dependent scattering at low doping and the progressive spectral broadening at larger $\delta$.

The analysis above establishes the division of labor between symmetry and correlations in the doped antiferromagnetic Hubbard model. Symmetry fixes when the polarization can be nonzero: whenever generalized particle--hole symmetry or the sublattice equivalence is preserved, exact compensation between spin channels is enforced at the level of the Kubo kernel and $P$ vanishes identically. Once both constraints are lifted, the correlations set everything quantitative: the magnitude of $P$, its non-monotonic doping dependence, and the location of the compensation point, which is fixed by the correlated hierarchy of scattering rates and the accompanying redistribution of spectral weight.

The local nature of the mechanism also clarifies its relation to recent DMFT studies of altermagnetic Hubbard models. Ref.~\cite{DelRe2025} solves the altermagnetic $t$--$t'$--$\delta$ model with an exact-diagonalization impurity solver and focuses on the spectral structure---Dirac points and their topology---and on the spin-resolved optical conductivity at zero field, where the spin asymmetry resides in the band dispersions. Here, by contrast, a structurally conventional antiferromagnet is solved with a CT-QMC impurity solver, and the observable is the dc spin-polarized transport switched on by doping and field, with the spin asymmetry residing in the local spectral functions and scattering rates. The two studies are thus complementary in model, method, and observables, while sharing the symmetry selection rule discussed in the main text.

\section{Comparison with the paramagnetic solution}
\label{SM:PM}

To isolate the role of the antiferromagnetic order in the transport polarization, we compare the N\'eel solution of the main text with the paramagnetic (PM) solution of the same model, obtained by enforcing sublattice equivalence, $\Sigma_{A\sigma}=\Sigma_{B\sigma}$, which reduces the two coupled impurity problems to a single field-polarized one. The comparison is carried out at the same $U=1.7$ and $T=0.05$, for dopings $0<\delta<\delta_c$ and the magnetic fields of Fig.~2 of the main text.

\begin{figure}[t]
\centering
\includegraphics[width=0.7\linewidth]{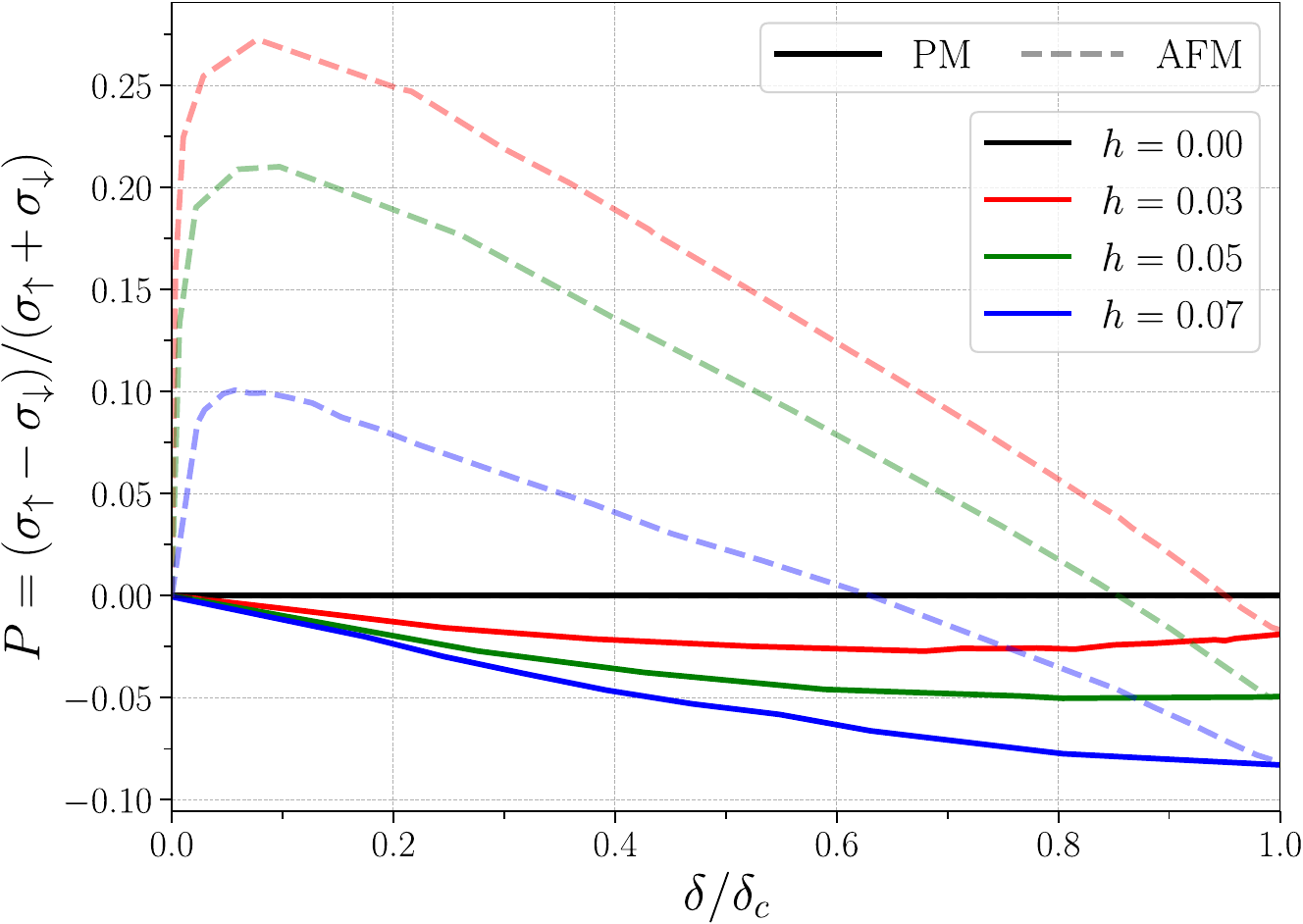}
\caption{Current polarization $P=(\sigma_\uparrow-\sigma_\downarrow)/(\sigma_\uparrow+\sigma_\downarrow)$ in the antiferromagnetic (dashed, faded) and paramagnetic (solid) solutions of the same model, at $U=1.7$ and $T=0.05$, for the magnetic fields of Fig.~2 of the main text. Dopings are normalized to the antiferromagnetic critical doping $\delta_c(h)$ of the corresponding field, the paramagnetic solution having no such boundary. Both solutions vanish identically at half filling, where the generalized particle--hole symmetry enforces $P \equiv 0$ irrespective of magnetic order, and at zero field. Away from half filling the paramagnetic polarization is negative at all dopings, monotonic, and without sign reversal, whereas the antiferromagnetic one is positive over most of the ordered range and up to an order of magnitude larger at low doping; the two converge as $\delta \to \delta_c$, where the staggered order collapses.}
\label{fig:SM_PM}
\end{figure}

Figure~\ref{fig:SM_PM} shows that the two solutions respond with opposite sign. Both vanish identically at half filling---as required by the generalized particle--hole symmetry, which is insensitive to the presence of N\'eel order---and at zero field. Away from half filling, the paramagnetic polarization is negative at every doping and varies smoothly and monotonically, saturating at large $\delta$, as expected for a field-polarized metal: it shows neither the low-doping enhancement nor the sign reversal of the ordered solution, and at low doping it is smaller by up to an order of magnitude.

The two curves converge as $\delta\to\delta_c$, where the staggered order collapses and the antiferromagnetic solution reduces to the paramagnetic one. This clarifies the origin of the compensation point: a negative, paramagnetic-like background is present at all dopings, while the positive, order-driven contribution is strongest where the N\'eel order is most robust and weakens with it. The sign reversal marks the doping at which the latter ceases to dominate the former---and hence why a stronger field, suppressing the staggered order, moves it to lower doping.

The effect reported here is therefore not the generic response of a doped metal in a Zeeman field. The magnitude of the polarization, its sign at low doping, and the existence of the compensation point are properties of the correlated antiferromagnetic state: the paramagnetic solution of the same model, at the same interaction strength, temperature, and field, yields only a weak response of opposite sign.

\bibliography{referencias}